\documentstyle[epsfig]{aipproc}

\def\one{{\sc i}}
\def\two{${\sc i\!i}$}
\begin{document}

   \title{A model for the high-energy emission\\ of Cyg X-1}
   \author{Igor V.~Moskalenko$^{\dagger\star}$, Werner Collmar$^\dagger$, 
       Volker Sch\"onfelder$^\dagger$}
   \address{
      $^\dagger$Max-Planck-Institut f\"ur extraterrestrische Physik,
      D-85740 Garching, Germany \\
      $^\star$Institute for Nuclear Physics,
      Moscow State University, 119 899 Moscow, Russia }
   \maketitle

\begin{abstract}
We construct a model of Cyg X-1 which describes self-consistently its
emission from soft X-rays to MeV $\gamma$-rays.  Instead of a compact
pair-dominated $\gamma$-ray emitting region, we consider a hot
optically thin and spatially extended proton-dominated cloud
surrounding the whole accretion disc. The $\gamma$-ray emission is due
to bremsstrahlung, Comptonization, and positron annihilation, while the
corona-disc model is retained for the X-ray emission.  We show that the
Cyg X-1 spectrum accumulated by {\sc osse}, {\sc batse}, and {\sc
comptel} in 1991--95, as well as the {\sc heao-3} $\gamma_1$ and
$\gamma_2$ spectra can be well fitted by our model (see
\cite{Moskalenko97} for details). The derived parameters are in
qualitative agreement with the picture in which the spectral changes
are governed by the mass flow rate in the accretion disc. In this
context, the hot outer corona could be treated as the
advection-dominated flow co-existing with a standard thin accretion
disc.

\end{abstract}

\section*{Introduction}

Cyg X-1 is believed to be powered by accretion through an accretion
disc. Its X-ray spectrum indicates the existence of a hot X-ray
emitting and a cold reflecting gas. The soft blackbody component is
thought to be thermal emission from an optically thick and cool
accretion disc \cite{Pringle81,Balucinska95}. The hard X-ray part
$\gtrsim10$ keV with a break at $\sim150$ keV has been attributed to
thermal disc emission Comptonized by a corona with a temperature of
$\sim100$ keV \cite{SunyaevTitarchuk80,LiangNolan84}.  A broad hump
peaking at $\sim20$ keV \cite{Done92}, an iron K$\alpha$ emission line
at $\sim6.2$ keV \cite{Barr85}, and a strong iron K-edge
\cite{Inoue89,Tanaka91} have been interpreted as signatures of Compton
reflection of hard X-rays off cold accreting material.  There have also
been reports of a hard component extending into the MeV region.  Most
famous was the so-called `MeV bump' observed at a $5\sigma$ level
during the {\sc heao-3} mission \cite{Ling87}. For a discussion of the
pre-{\sc cgro} data see \cite{OwensMcConnell92}.

The average X-ray flux of Cyg X-1 shows a two-modal behavior. Most of
its time it spends in a so-called `low' state where the soft X-ray
luminosity is low.  There are occasional periods of `high' state
emission. Remarkable is the anticorrelation between the soft and hard
X-ray components \cite{LiangNolan84}, which is clearly seen during
transitions between the two states.

Since its launch in 1991 {\sc cgro} has observed Cyg X-1 several
times.  The {\sc comptel} spectrum shows significant emission out to
several MeV, which however, remained always more than an order of
magnitude below the MeV bump reported by {\sc heao-3}.  The spectrum
accumulated between '91 and '95 by the {\sc comptel}
\cite{McConnell97}, {\sc batse} \cite{Ling97}, and {\sc osse}
\cite{Phlips96} instruments is shown in Fig.~\ref{fig2}.  Although the
{\sc osse} and {\sc batse} normalizations are different, the spectral
shape is very similar. Table 1 shows the average luminosity of Cyg X-1
(at 2.5 kpc) as derived from the {\sc batse-comptel} spectrum.

\begin{figure}[t!]
   \begin{picture}(140,52)(0,0)
      \put(70,5){ \makebox(70,50)[tc]{%
         \psfig{file=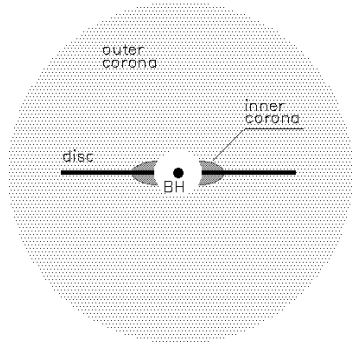,height=55mm,clip=}}} 
      \put(75,0){ \begin{minipage}[bc]{60mm}
         \caption[]{A schematic view.} 
         \label{fig1}
      \end{minipage}}

      \put(5,15){ \parbox[b]{60mm}{ \footnotesize
         {\bf TABLE 1.}\ Luminosity of Cyg X-1.
         \setcounter{table}{1}
         \begin{tabular}{lc} 
            \tableline 
            \noalign{\medskip}
Energy band & Luminosity, $10^{36}$ erg/s \\
            \hline
$\geq0.02$ MeV  & $26$ \\
0.02--0.2 MeV   & $20.5$ \\
0.2--1 MeV      & $4.8$\\
$\geq1$ MeV     & $0.6$\\
          \tableline
      \end{tabular}}}
   \end{picture}
\end{figure}

\section*{The Model}
The existence of a compact pair-dominated core around a {\sc bh} in Cyg
X-1 is probably ruled out by the {\sc cgro} observations.  A signature
of such a core would be a bump \cite{LiangDermer88,Liang90} similar to
the one reported by {\sc heao-3}. However no evidence for such a bump
was found in the {\sc cgro} data \cite{McConnell97,Phlips96}.
Additionally, the luminosity of Cyg X-1 above 0.5 MeV, though small,
exceeds substantially the Eddington luminosity for pairs, which is
$\sim2000$ times lower than for a hydrogen plasma.  Also the hard MeV
tail can not be explained by Comptonization in a corona ($kT\sim100$
keV) and thus another mechanism is required.

We consider the proton-dominated optically thin solution
\cite{Svensson84}, $\Theta\equiv\frac{kT}{m_ec^2}\lesssim1$, where the
$\gamma$-ray emission is attributed to a spatially extended cloud
surrounding the whole accretion disc (Fig.~\ref{fig1}), the outer
corona, which emits via bremsstrahlung, Comptonization, and positron
annihilation, and analyze possible consequences of that.  We adopt a
standard model for X-rays, where the hard X-rays are produced by
Comptonization of the soft X-ray emission in an inner corona, and the
soft X-rays are a composition of the local blackbody emission from the
disc and the reflection component. The optical depth of the outer
corona has to be so small that the disc and inner corona emission is
only slightly reprocessed in it.

The spectral modelling has been carried out with the $ee$- and
$e^+e^-$-brems\-strahl\-ung emissivities given by numerical fits of
\cite{StepneyGuilbert83,Haug87}.  For the $ep$-bremssrahlung and
annihilation emissivities we use the integration formulas of
\cite{StepneyGuilbert83,Dermer84}.  To calculate the effect of Compton
scattering we follow the model of \cite{HuaTitarchuk95}, which
generally agrees well with Monte Carlo simulations except at high
temperatures, $\Theta\sim1$, and small optical depth. But it still
provides a correct spectral index. We found that a power-law with a
cutoff, $\propto\left[\frac{E_0}E\right]^{\alpha+1} (1-e^{-kT/E})$,
where $\alpha$ is determined by the equation of
\cite{TitarchukLyubarskij95}, gives a reasonable agreement with Monte
Carlo simulations up to $\sim3$ MeV. The chosen normalization provides
a correct value of the amplification factor \cite{Dermer91}.  The
emission of the accretion disc which is reprocessed by the inner corona
was taken monoenergetic, $E_0=1.6kT_{bb}$, where $kT_{bb}\approx0.13$
keV is the effective temperature of the soft excess
\cite{Balucinska95}.  The intensity of the narrow annihilation line
from the disc plane can be estimated by $I_a\simeq\frac{n_+c}{4}
\frac{R_d^2}{D^2} \cos i_d$, where $n_+$ is the $e^+$ number density in
the outer corona, $c$ the speed of light, $R_d$ the disc radius, $D$
the distance, and $i_d$ ($\approx40^\circ$) the inclination angle of
the disc plane.


The fitting parameters are: $kT_i$, $\tau_i$, and $kT_o$, $\tau_o$, the
temperature and optical depth of the inner ({\it i}) and outer ({\it
o}) coronae (spheres), $L^*_{\rm soft}$, the luminosity of the {\it
disc} effectively Comptonized by the {\it i}-corona, $L_{\rm soft}$,
the {\it total} effective luminosity of the central source in soft
X-rays illuminating the {\it o}-corona, $R$, the {\it o}-corona radius,
and, $Z=n_+/n_p$, the positron-to-proton ratio in it.

The bremsstrahlung and annihilation photon fluxes from the outer corona
are proportional to $R^3n_p^2$.  Thus, if the annihilation contributes
significantly, there is a continuum of solutions given by an equation
$R^3n_p^2 Z(1+Z)=\frac{R\tau_o^2 Z(1+Z)} {\sigma_T^2(1+2Z)^2}=const$,
at $kT_o, \tau_o$ fixed, where $\frac{Z(1+Z)}{(1+2Z)^2}$ varies slowly
for $Z\gtrsim0.5$.  For a negligible positron fraction the continuum of
solutions is defined by $\tau_o = R n_p \sigma_T = const$, where $R\leq
R_{\rm max}$, and $R_{\rm max}$ is fixed from fitting.

\begin{figure}[t!]
   \begin{picture}(140,35)(9,-3)
      \put(0,0){\makebox(45,35)[tl]{%
         \psfig{file=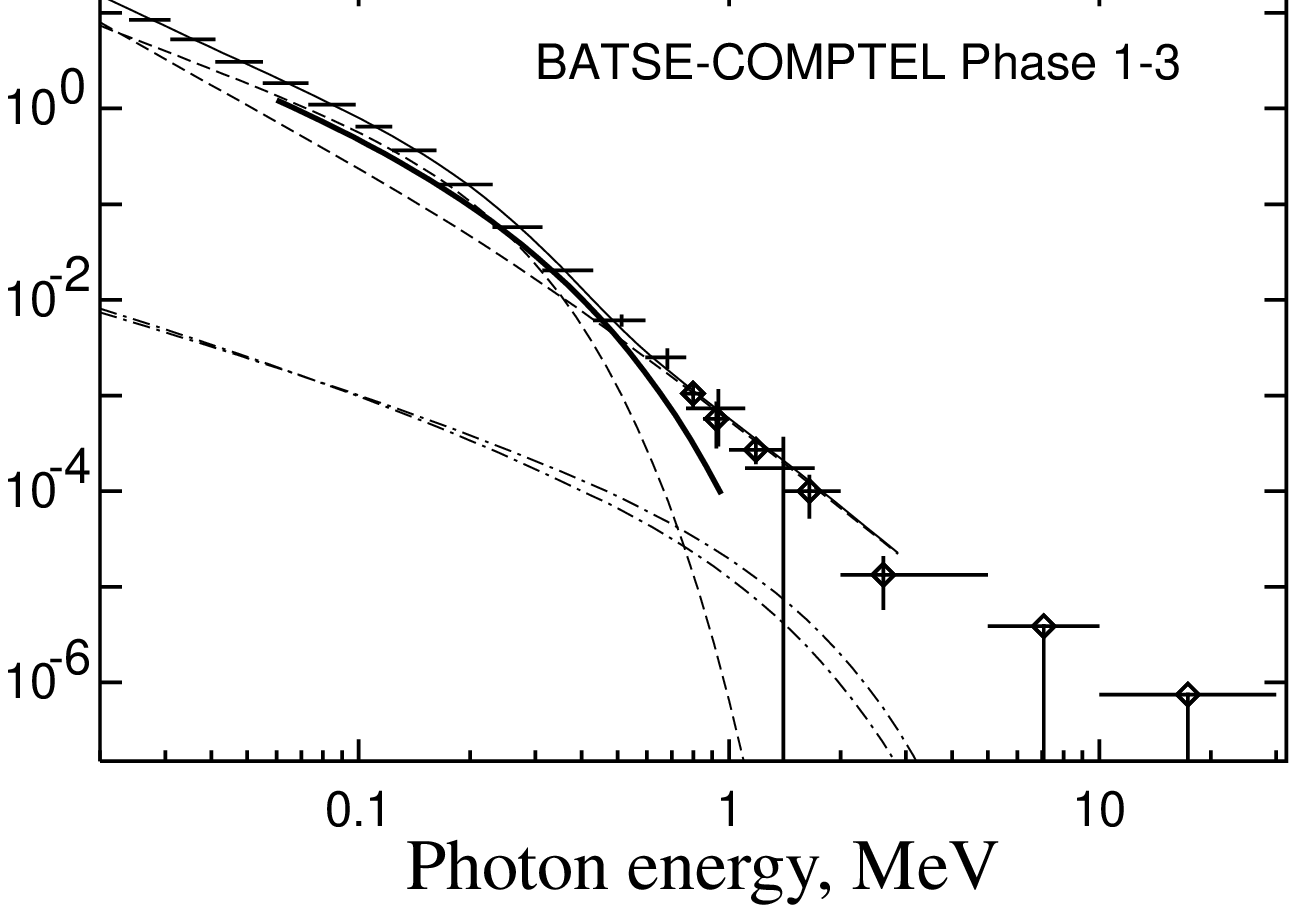,height=100mm,width=73mm,clip=}}}
      \put(46,0){\makebox(45,35)[tl]{%
         \psfig{file=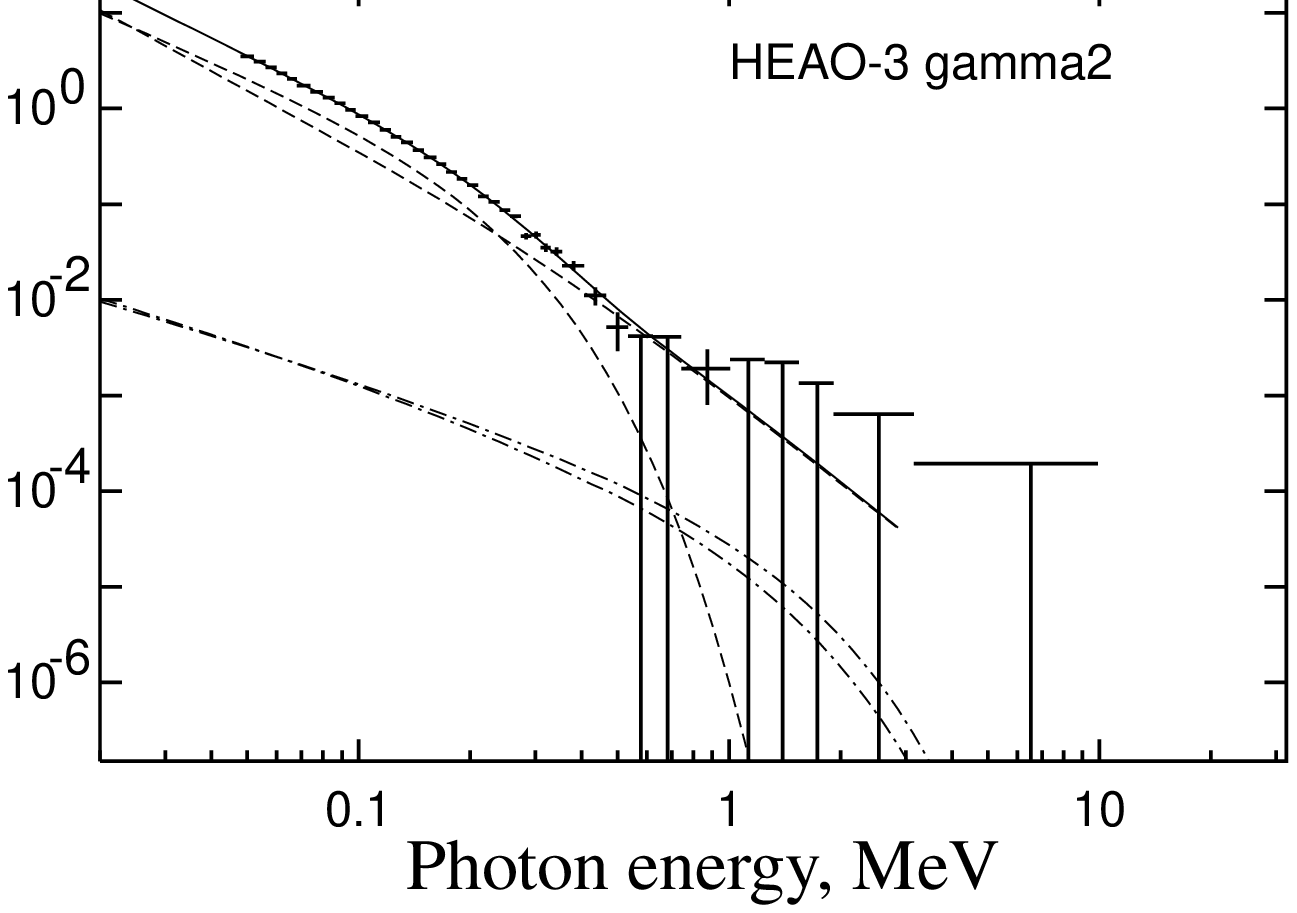,height=100mm,width=73mm,clip=}}}
      \put(92,0){\makebox(45,35)[tl]{%
         \psfig{file=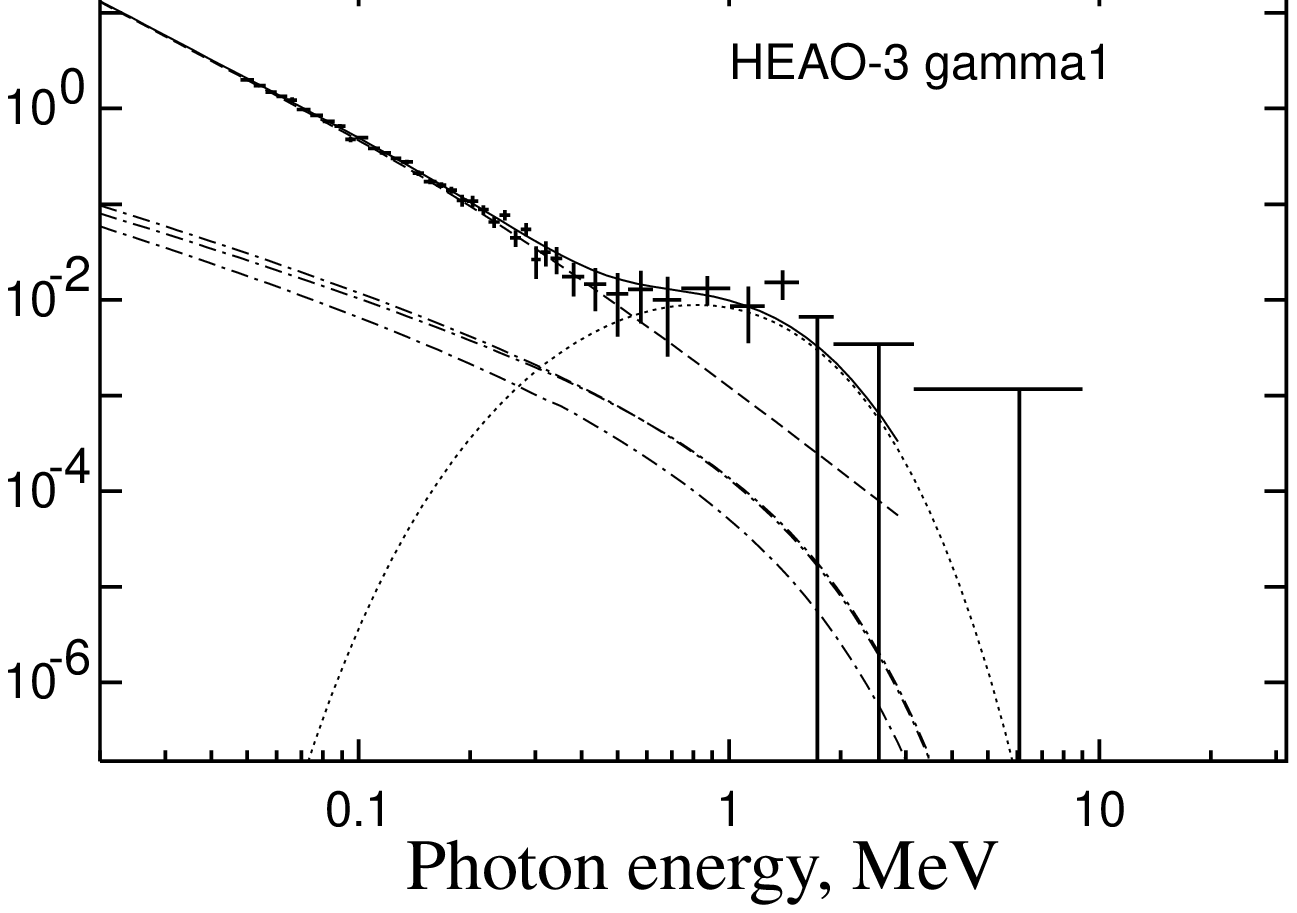,height=100mm,width=73mm,clip=}}}
   \end{picture}
\caption[]{%
{\it Left panel:} the Cyg X-1 spectrum based on the {\sc cgro} Phase
1--3 observations ($\Diamond$: the {\sc comptel} data
\cite{McConnell97}, +: the {\sc batse} data \cite{Ling97}). The thick
solid line is the best fit to the time average {\sc osse} spectrum
\cite{Phlips96}. {\it Central and right panels:} the {\sc heao-3}
$\gamma_2$, $\gamma_1$ spectra \cite{Ling87}.  In all panels the thin
solid lines represent our model fit for the parameter sets \one. The
spectral components shown are the annihilation line (dotted line),
$ee$-, $e^+e^-$-, $ep$-bremsstrahlung (dash-dot), dashed lines: the
Comptonized spectra from the {\it i}- and {\it o}-corona (shown up to 3
MeV, where it agrees with simulations, and also significant data points
are available).}
\label{fig2}

\end{figure}

\section*{Results and Discussion}
\begin{table}[t!]
   \caption[]{The best fit model parameters.}
   \label{tab2} 
   \begin{tabular}{lcclcclcc}
& \multicolumn{2}{c}{{\sc cgro} Phase 1--3} &&
\multicolumn{2}{c}{{\sc heao-3:} $\gamma_2$-state} &&
\multicolumn{2}{c}{$\gamma_1$-state}  \\
      \cline{2-3}\cline{5-6}\cline{8-9}
\raisebox{2ex}[0pt]{Parameters} & \one & \two && \one & \two &&
   \one\tablenote{The inner corona is small or even absent at all;\hfill
   $^{\rm b}$For $R$, $n_p$, $Z$ dependence see text.} 
   & \two \\
      \hline
Soft X-ray luminosity, $L_{\rm soft}$ ($10^{36}$ erg s$^{-1}$)
                         & 9.7  & 10.6 && 10.6& 10.7  && 9.8  & 7.9\\
{\it i}-corona temperature, $kT_i$ (keV) 
                         & 75.9 & 73.9 && 95  & 94.9  &&\bf --& 93.0\\
{\it i}-corona optical depth, $\tau_i$   
                         & 2.30 & 2.40 && 1.41 & 1.42 &&\bf --& 1.44\\
$L^*_{\rm soft}$, $10^{36}$ erg s$^{-1}$
                         & 0.86 & 0.83 && 1.96 & 1.95 &&\bf --& 0.51\\
{\it o}-corona temperature, $kT_o$ (keV)
                         & 430  & 479  && 450  & 448  && 346  & 361 \\
{\it o}-corona optical depth, $\tau_o$
                         & 0.05 & 0.037&& 0.056& 0.056&& 0.12 & 0.10\\
{\it o}-corona radius, $R$ ($10^8$ cm)$^{\rm b}$
       & $\lesssim100$ & $\lesssim100$ && 
                                  $\lesssim100$ & 150 && 391 & 812 \\
Positron-to-proton ratio, $Z^{\rm b}$
                         & 0    & 0    && 0    & 1.00 && 1.0  & 0.5\\
Proton number density, $n_p$ ($10^{10}$ cm$^{-3}$)$^{\rm b}$
                         &\bf --&\bf --&&\bf --& 187  && 154  & 93 \\
Accretion disc radius, $R_d$ ($10^8$ cm)
                         &\bf --&\bf --&&\bf --& 1    && 1    & 1  \\
$I_a$, $10^{-5}$ photons cm$^{-2}$ s$^{-1}$
                         & 0    & 0    && 0    & 0.18 && 0.15 & 0.04\\
$\chi^2_\nu$             & 3.1  & 3.1  && 1.4  & 1.4  && 0.9  & 0.9 \\
   \end{tabular}
\end{table}

The observed spectra of Cyg X-1 are shown in Fig.~\ref{fig2} together
with our model calculations. The parameters obtained from spectral
fitting are listed in Table \ref{tab2}. For comparison two parameter
sets with the same $\chi^2_\nu$ are shown, however the first one (\one)
seems to be more physical.

The average {\sc batse-comptel} spectrum corresponds probably to the
normal state of Cyg X-1.  Only two components contribute: the
Comptonized emission from the inner and outer coronae.  The parameters
obtained for the {\sc heao-3} $\gamma_2$ state are similar, though the
upper limits at $E_\gamma\gtrsim1$ MeV allow some positron fraction
(\two).  For the {\sc heao-3} $\gamma_1$ `bump' spectrum the outer
corona size is several times larger, while the inner corona is small or
even absent at all (set \one). The non-negligible positron fraction
(for $R,n_p,Z$ dependence see above) is too high to be produced in the
optically thin outer corona \cite{Svensson84}.  Therefore, we suggest a
positron production mechanism, which could sometimes operate in the
inner disc.  The radiation pressure would necessarily cause a pair
wind, which serves as energy input into the {\it o}-corona thereby
increasing its radius.

The small luminosity of the disc which is Comptonized by the inner
corona, $L^*_{\rm soft}\approx10^{36}$ erg/s, probably implies a
geometry where only the inner part of the disc is effectively covered
by the corona.  Otherwise, if the corona forms a disc-like structure
where the intensity depends on the inclination angle, then it should
cover almost all of the X-ray emitting area of the disc.

The soft ($<10$ keV) X-ray luminosity of Cyg X-1 is
$\sim8.5\times10^{36}$ erg/s on average \cite{LiangNolan84}, while
during the {\sc heao-3} $\gamma_1$, $\gamma_2$ states it was even lower
\cite{Ling87}.  Taking into account that for hard X-ray photons the
Comptonization efficiency in the hot plasma drops substantially
\cite{HuaTitarchuk95} while the number of photons decreases as well,
the obtained values, $L_{\rm soft}\approx10^{37}$ erg/s, match the
data.

No pairs are required to reproduce the spectrum of Cyg X-1 in its
normal state.  If one takes $I_a\approx4.4\times10^{-4}$ photons
cm$^{-2}$ s$^{-1}$ \cite{LingWheaton89} for the annihilation line flux
in the $\gamma_1$ state, it allows the accretion disc radius to be
estimated to $R_d\sim1.7\times10^9$ cm (set \one). The allowed upper
limit derived from optical measurements is $R_d\approx6\times10^9$ cm
$(M/10M_\odot)$ \cite{LiangNolan84}.

The obtained parameters are consistent with a picture where the
spectral changes are governed by the mass accretion rate $\dot{M}$
\cite{Narayan96}. The $\gamma_1$ state probably corresponds to a
smaller $\dot{M}$ compared to the normal state.  In this context, the
$\gamma$-ray luminosity should anticorrelate with the hard X-ray
luminosity.  The extended hot outer corona can be treated as the
advection-dominated accretion flow (ADAF) co-existing with a cool
optically thick disc, though in contrast to a standard ADAF
\cite{Narayan96}, the electrons here are hot and the protons are cold.
This is possible since cooling via bremsstrahlung and Coulomb
$ep$-collisions at low density is unimportant while small optical depth
prevents from effective Compton cooling.  The adequate heating could be
provided by the electron thermal conduction from a region with nearly
virial ion temperature.

\footnotesize
Useful discussions with N.Shakura, R.Narayan, L.Titarchuk, and
M.Gilfanov are greatly acknowledged. We are particularly grateful to
M.McConnell for providing us with the combined spectra of Cyg X-1 prior
to publication, and E.Churazov for Monte Carlo simulations of
Comptonization in $\Theta\sim1$, $\tau\approx0.1-0.05$ plasma.

\renewcommand{\baselinestretch}{0.91} 

\normalsize

\begin{references}
\bibitem{Moskalenko97} Moskalenko I.\ V., Collmar W., Sch\"onfelder V.,
   {\it ApJ}, submitted (1997)
\bibitem{Pringle81} Pringle J.\ E., {\it ARA\&A} {\bf 19}, 137 (1981) 
\bibitem{Balucinska95} Ba{\l}uci\'nska-Church M., et al., 
   {\it A\&A} {\bf 302}, L5 (1995) 
\bibitem{SunyaevTitarchuk80} Sunyaev R.\ A., Titarchuk L.\ G., 
   {\it A\&A} {\bf 86}, 121 (1980)
\bibitem{LiangNolan84} Liang E.\ P., Nolan P.\ L., 
   {\it Spa.\ Sci.\ Rev.}\ {\bf 38}, 353 (1984)
\bibitem{Done92} Done C., et al., 
   {\it ApJ} {\bf 395}, 275 (1992)
\bibitem{Barr85} Barr P., White N., Page C.\ G., {\it MNRAS} {\bf 216}, 
   65p (1985)
\bibitem{Inoue89} Inoue H., in {\it Proc.\ 23rd ESLAB Symp.},
   Noordwijk: ESA, 1989, {\bf 2}, p.783
\bibitem{Tanaka91} Tanaka Y., {\it Lect.\ Notes in Phys.}\ {\bf 385}, 98 (1991)
\bibitem{Ling87} Ling J.\ C., et al.,
   {\it ApJ} {\bf 321}, L117 (1987)
\bibitem{OwensMcConnell92} Owens A., McConnell M.\ L., {\it Comments
   Astrophys.}\ {\bf 16}, 205 (1992)
\bibitem{McConnell97} McConnell M., et al., in {\it AIP Proc.}, 1997,
   this Symposium
\bibitem{Ling97} Ling J.\ C., et al., {\it ApJS}, submitted (1997)
\bibitem{Phlips96} Phlips B.\ F., et al., {\it ApJ} {\bf 465}, 907 (1996)
\bibitem{LiangDermer88} Liang E.\ P., Dermer C.\ D., {\it ApJ} {\bf 325},
   L39 (1988)
\bibitem{Liang90} Liang E.\ P., {\it A\&A} {\bf 227}, 447 (1990)
\bibitem{Svensson84} Svensson R., {\it MNRAS} {\bf 209}, 175 (1984)
\bibitem{StepneyGuilbert83} Stepney S., Guilbert P.\ W., {\it MNRAS} {\bf 204}, 
   1269 (1983)
\bibitem{Haug87} Haug E., {\it A\&A} {\bf 178}, 292 (1987)
\bibitem{Dermer84} Dermer C.\ D., {\it ApJ} {\bf 280}, 328 (1984)
\bibitem{HuaTitarchuk95} Hua X.-M., Titarchuk L., {\it ApJ} {\bf 449},
   188 (1995)
\bibitem{TitarchukLyubarskij95} Titarchuk L., Lyubarskij Yu., 
   {\it ApJ} {\bf 450}, 876 (1995)
\bibitem{Dermer91} Dermer C.\ D., Liang E.\ P., Canfield E., {\it ApJ} 
   {\bf 369}, 410 (1991)
\bibitem{LingWheaton89} Ling J.\ C., Wheaton Wm.\ A., {\it ApJ} {\bf 343},
   L57 (1989)
\bibitem{Narayan96} Narayan R., {\it ApJ} {\bf 462}, 136 (1996)

\end{references}
\end{document}